\documentstyle[aps,pra,amsmath,amssymb,citesort,epsfig]{revtex}

\begin{document}

\title{Homogeneous nucleation of a non-critical phase
near a continuous phase transition}
\author{{\bf Richard P. Sear}\\
~\\
Department of Physics, University of Surrey\\
Guildford, Surrey GU2 7XH, United Kingdom\\
email: r.sear@surrey.ac.uk}

\maketitle

\begin{abstract}
Homogeneous nucleation of a new phase near a second,
continuous, transition, is considered. The continuous transition is
in the metastable region associated with the first-order phase transition,
one of whose coexisting phases is nucleating.
Mean-field calculations show that
as the continuous transition is approached, the size of the
nucleus varies as the response function of the order parameter
of the continuous transition.
This response function diverges at the continuous transition, as does
the temperature derivative of the free energy
barrier to nucleation. This rapid drop of the barrier as the
continuous transition is approached means that the continuous
transition acts to reduce the barrier to nucleation at
the first-order transition. This may be useful in the
crystallisation of globular proteins.
\end{abstract}

\section{Introduction}

When a phase transition is first order the formation of a new phase
is an activated process \cite{note}. A nucleus of the new phase must form,
and overcome a free-energy barrier before it can grow into a new
phase. The rate at which nuclei overcome a barrier of height
$\Delta F^*$ scales as $\exp(-\Delta F^*/kT)$ and so this rate
is very sensitive to the barrier height \cite{debenedetti}.
For definiteness consider a first-order phase transition at a temperature
$T_{\alpha}$ in which the low temperature phase is the ordered
phase, and the high temperature phase is the disordered phase.
If we cool the disordered phase below $T_{\alpha}$ but the
barrier to nucleation of the ordered phase
is very high then the rate at which nuclei of the
ordered phase form will be
effectively zero and the ordered phase will not form even though its
free energy is lower than that of the disordered phase.
The disordered phase will persist; it is said
to be metastable. Here we calculate $\Delta F^*$ for nucleation
near a second, continuous, transition, which we call transition $\beta$. 
Continuous transitions are critical points and so
exhibit universal and beautiful behaviour,
the thermodynamic and correlation functions contain power law terms
with exponents which depend only on dimensionality and the
symmetry of the order parameter \cite{kadanoff,chaikin}.
A priori, we might expect that $\Delta F^*$ might also contain
a power law term with an exponent which depends only on dimensionality
and the symmetry of the order parameter.
This would allow us to make predictions about how $\Delta F^*$
varied near a critical point which would apply to a whole
class of systems.
Below, we present the results of calculations within mean-field theory,
for an Ising-like continuous transition in three dimensions.
We determine the singular power law
part of the free energy barrier $\Delta F^*$:
just above $T_{\beta}$,
it varies as $\xi^{-1}$, where $\xi$ is the correlation length
associated with the order parameter of transition $\beta$.
This singular part means that the derivative of $\Delta F^*$ with
respect to temperature diverges as the critical point is approached:
the barrier to nucleation drops rapidly just above the transition.
This agrees with the pioneering simulations of ten Wolde and Frenkel
\cite{tenwolde97} who found an anomalously low $\Delta F^*$ near
the critical point of a metastable fluid-fluid transition.

As far as the author is aware,
nucleation near a critical point has only been considered for
a fluid-fluid critical point within a strongly first-order fluid-crystal
transition. This work was
inspired by the observation that globular proteins often crystallise
at temperatures close to where we might expect a metastable
fluid-fluid critical point \cite{george94,rosenbaum96}, and that
the phase diagrams of some globular proteins do possess a
metastable fluid-fluid transition \cite{broide91,muschol97}.
But as the effect we will examine is due to the
decreasing free energy cost of fluctuations near a critical point it
is universal, i.e., applies to any other system in the same
universality class, that of the three-dimensional Ising model.
Indeed, it is easy to show that it also applies to systems in
the universality classes of the Ising model in other dimensionalities
\cite{unpub}.
See Refs. \cite{talanquer98,haas00,dixit00,pini00,search}
and references therein for recent work.

The next section briefly sets out the standard Landau theory for
a continuous transition with a scalar order parameter. Section III
then calculates the free energy of a nucleus within a simple mean-field
theory of the square-gradient type. Derivatives with respect to
temperature and external field are also found. The final section is
a conclusion. See Refs. \cite{kadanoff,chaikin} for an introduction
to systems near critical points and Ref. \cite{debenedetti} for an
introduction to homogeneous nucleation.

\section{Bulk behaviour}

We have a system, which at equilibrium has one phase transition:
a strongly first order transition, transition $\alpha$, which
is at a temperature $T_{\alpha}$. For definiteness we let the
high-temperature phase be
the disordered phase and the low-temperature phase be the ordered phase.
If we consider a very pure sample \cite{note} then
we can supercool down to temperatures below $T_{\alpha}$ to obtain metastable
states \cite{debenedetti}, i.e., the disordered phase is stable
for long (with respect to the relaxation time of the system)
periods of time
over some temperature range just below $T_{\alpha}$.
It is stable because the formation of the ordered phase is an activated
process, the ordered phase
must nucleate, overcoming some free energy barrier $\Delta F^*$
which will be a strong function of temperature and which diverges as
$T\rightarrow T_{\alpha}^-$. Here, we are interested in how $\Delta F^*$
behaves if as we cool down in the metastable disordered phase,
we approach a continuous
transition, transition $\beta$, at some
temperature $T_{\beta}<T_{\alpha}$.

We will assume that the nucleation barrier to transition $\alpha$
is so large that
it is possible to start from some temperature $T>T_{\alpha}$ and slowly
cool down past $T_{\alpha}$ either to and below $T_{\beta}$,
or at least to a temperature only a little above $T_{\beta}$, without
transition $\alpha$ occurring. If it is
possible to cool slowly down to $T_{\beta}$,
then transition $\beta$ is said to be metastable
\cite{debenedetti}: it is observable.
If it is not possible to cool slowly down to $T_{\beta}$
without transition $\alpha$ occurring then clearly transition $\beta$ is
not observable; it is unstable not metastable
with respect to transition $\alpha$ \cite{debenedetti}.
We will be studying nucleation of the low temperature phase
of transition $\alpha$ as $T_{\beta}$ is approached from above
and so we will not only be determining
the effect of transition $\beta$ on $\alpha$ but also looking at whether
or not $\beta$ is observable. Roughly speaking, if the proximity of
a transition $\beta$ acts to strongly enhance the nucleation rate of
transition $\alpha$, then this nucleation rate may become large thus
rendering transition $\beta$ unobservable.
We should also mention that we are using temperature as
our variable simply for definiteness, we could replace it by another field
variable, e.g., pressure, without changing our conclusions.
  
\begin{figure}
\begin{center}
\caption{
\lineskip 2pt
\lineskiplimit 2pt
Schematic of a nucleus of the ordered phase of transition $\alpha$
near transition $\beta$. The core of the ordered phase of transition
$\alpha$ is solid black, and the perturbation this causes in the
surroundings is the shaded circle of radius the correlation length $\xi$.
The sphere with
radius $r_c$ which divides the nucleus
into a core and surroundings is denoted by a dashed circle.
\label{figschem}
}
\vspace*{0.2in}
\epsfig{file=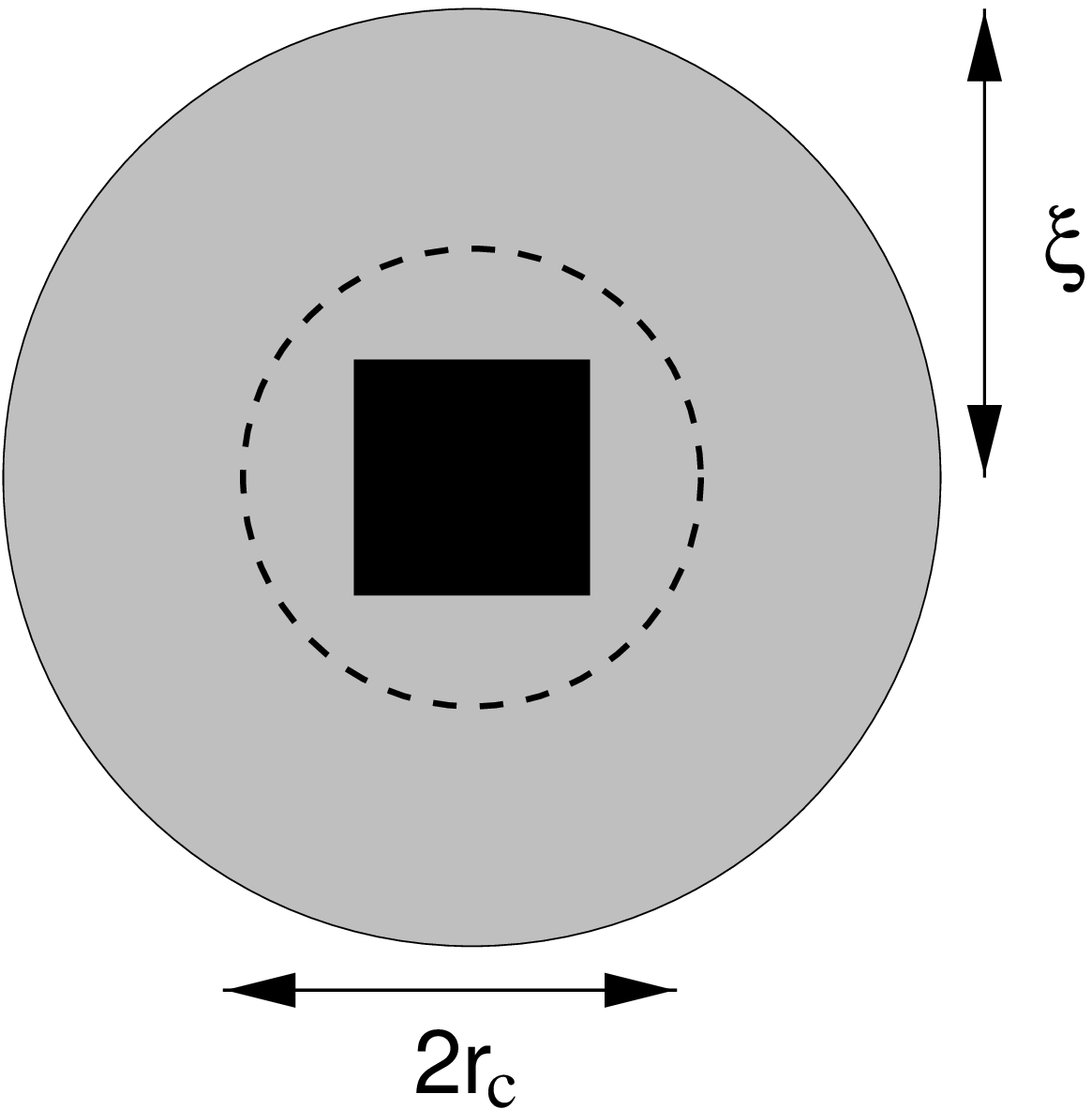,width=2.5in}
\end{center}
\end{figure}

So, starting from high temperature and cooling down below $T_{\alpha}$
we can approach $T_{\beta}$. The order parameter of transition $\beta$ is
denoted by $m$ and it may or may not be related to that of transition $\alpha$.
The external field which couples to $m$ is $h$.
The theory here will be mean-field in nature but rather general.
We only have to assume that the nucleus of the ordered phase
of transition $\alpha$ 
has a core which has properties close to that of the bulk ordered phase
(c.f., the assumptions which underly classical nucleation
theory \cite{debenedetti})
and that this core couples to the order parameter of transition $\beta$.
By coupling to $m$ we mean that if there is a nucleus at the origin, then
the local value of $m$, $m_r(r)\ne m$, where $r$ is the distance
from the centre of the nucleus.
Both assumptions are very reasonable:
for a strongly first-order phase transition it is difficult
to imagine a situation where the nucleus does not have a core with near
bulk properties, and the core of the nucleus must perturb its surroundings
and so, in the absence of special symmetries, will locally perturb the order
parameter of transition $\beta$. Figure \ref{figschem} is a schematic
of the nucleus.

Near $T_{\beta}$ we use a Landau theory for the transition $\beta$.
The Landau theory of a continuous transition
is simple, it is a textbook problem, see for example Chaikin and
Lubensky's \cite{chaikin} or Kadanoff's \cite{kadanoff}. The
bulk free energy per unit volume $f(m)$ as a function of the order
parameter $m$ is
\begin{equation}
f(m)=\frac{1}{2}a(T-T_{\beta})m^2+m^4-hm,
\label{landau}
\end{equation}
The transition is at $T_{\beta}$ at $h=0$. We will only examine behaviour
at $h=0$ but we retain $h$ in order to look at the response of the system
to an external field which couples to $m$.
Below, when we study the nucleus near transition $\beta$,
we will find that the outermost part of the density profile of the nucleus is
controlled by the response function of $m$, $\chi$, defined by
\begin{equation}
\chi^{-1}=\left(\frac{\partial h}{\partial m}\right)=
\left(\frac{\partial^2 f}{\partial m^2}\right),
\label{chidef}
\end{equation}
which is, using Eq. (\ref{landau}),
\begin{equation}
\chi^{-1}= a(T-T_{\beta})  ~~~~~~ T > T_{\beta}.
\label{chic}
\end{equation}

\section{The nucleus}

We split the nucleus into two parts: the core and the fringe. The core is
that part of the nucleus less than $r_c$ from its centre and the fringe
is that part farther away than $r_c$. The fringe of the
nucleus is assumed to be
spherically symmetric. The core of the nucleus contains at its centre
a volume which is close to the bulk ordered phase of transition $\alpha$.
The fringe is essentially the region which surrounds this core and is
perturbed by the core. Its radius is therefore the correlation length $\xi$
of $m$ and so diverges as $T_{\beta}$ is approached. As we are concentrating
on universal aspects of the nucleus and of $\Delta F^*$ we will replace the
core by a boundary condition on $m_r$ in the fringe. We set
$m_r(r=r_c)=m_c$ which is taken to be independent of temperature and
of $h$.
Also, $m_r(r\rightarrow\infty)=m$, which is just the obvious
boundary condition that $m$ must tend towards its bulk value
far from the nucleus. Note that as we are above the transition and
are working at zero field, $m=0$ but
we will retain an explicit $m$ dependence in order to be able to take
derivatives of $\Delta F^*$.
In the fringe and near $T_{\beta}$,
we need only consider the order parameter for transition $\beta$ and
the variations in $m_r$ will not be large. Therefore, we employ a
standard Landau-Ginsburg or square-gradient
functional for the excess free energy $\Delta F$ of an
inhomogeneity in an otherwise homogeneous phase,
\cite{cahn58,cahn59,widom85,debenedetti}
\begin{equation}
\Delta F  = \Delta F_c + \int_{r\ge r_c}\left[
\Delta f(m_r) + \kappa\left(\nabla m_r\right)^2 \right]
{\rm d}{\bf r},
\label{cahn}
\end{equation}
where
\begin{equation}
\Delta f (m_r)=f(m_r)-f(m)-h(m_r-m),
\label{omega}
\end{equation}
is the bulk free energy change per unit volume to go from $m$ to $m_r$.
The excess free energy $\Delta F$ is the free energy with a nucleus minus that
without a nucleus; $\Delta F_c$ is the contribution of the core.
The second term
within the brackets of Eq. (\ref{cahn}) is a gradient term: the
free energy cost due to variations in space of $m_r$. It is proportional to
the gradient squared which is the lowest order term in a gradient
expansion and so is only adequate when $m$ is slowly
varying. The coefficient, $\kappa$, of this term is
taken to be a constant.
The total excess of $m$ due to a nucleus is equal to the
integrated value of $m_r-m$. Defining $\Delta m({\bf r})=m_r({\bf r})-m$,
then the integral over all space of this function gives the total
excess of the order parameter due to the nucleus,
\begin{equation}
\Delta m= \int \Delta m({\bf r}) {\rm d}{\bf r}.
\label{nstardef}
\end{equation}

The free energy barrier $\Delta F^*$ is the value of $\Delta F$ for
the nucleus when it is at its maximum, at the top of the barrier.
The nucleus at the top of the barrier is called the critical nucleus
\cite{critnote}.
For the
critical nucleus we may set the functional derivative of
$\Delta F$ with respect to the profile $m_r({\bf r})$
to zero,
\begin{eqnarray}
\left(\frac{\partial \Delta f(m_r) }{\partial m_r}\right)-2\kappa\nabla^2 m_r
= 0 ~~~~r>r_c.
\label{max}
\end{eqnarray}
Once we have solved Eq. (\ref{max}) we can insert the solution
into Eq. (\ref{cahn}) to obtain the excess free energy of the
critical nucleus, $\Delta F ^*$. 

The fringe is the outermost part of the nucleus, where 
$m_r$ is near the bulk value $m$. So
we can use a Taylor expansion about $m_r=m$ for $\Delta f$,
\begin{eqnarray}
\Delta f(m_r)
&=&\frac{1}{2}\chi^{-1}\left(\Delta m\right)^2+\cdots
\label{taylor1}\\
\left(\frac{\partial \Delta f(m_r)}{\partial m}\right)&=&
\chi^{-1}\Delta m+\cdots,
\label{taylor}
\end{eqnarray}
because both $\Delta f$ and its
first derivative are zero for $m_r=m$,
and the second derivative is $\chi^{-1}$ [Eq. (\ref{chidef})].
Substituting Eq. (\ref{taylor}) into Eq. (\ref{max}) we have
\begin{equation}
\chi^{-1}\Delta m(r)-
2\kappa\nabla^2 \Delta m(r) =0,
\label{helm}
\end{equation}
which has a solution of the Ornstein-Zernike form,
\begin{equation}
\Delta m(r) = \left(m_c-m\right)
\left(\frac{r_c}{r}\right)\exp\left[(r_c-r)/\xi\right],
\label{oz}
\end{equation}
with $\xi$ the correlation length for $m$, given by
\begin{eqnarray}
\xi &=&\left(2\kappa\chi\right)^{1/2}
\label{xia}
\\ \xi &=&
\left(2\kappa/a\right)^{1/2}\left(T-T_{\beta}\right)^{-1/2}
~~~~~~ T > T_{\beta},
\label{xi}
\end{eqnarray}
where Eq. (\ref{xia}) defines $\xi$ and we used Eq. (\ref{chic})
to obtain expressions for $\xi$ near $T_{\beta}$, Eq. (\ref{xi}).
To obtain Eq. (\ref{oz}) the boundary conditions
$m_r(r\rightarrow\infty)=m$ and $m(r_c)=m_c$ were employed.
It is not necessary to specify
$r_c$ or $m_c$ beyond saying that they should be such that
$m_c-m$ is small and so $\Delta m(r)$ will, as required
for Eq. (\ref{helm}), be small for $r\ge r_c$.
From Eq. (\ref{oz}) we see that
the width of the fringe is, as we expected,
of the order of the correlation length $\xi$ for $m$.

Having obtained the density profile, Eq. (\ref{oz}), we can substitute
this into Eq. (\ref{cahn}), using Eq. (\ref{taylor1}) for $\Delta f$, and
obtain an expression for the free energy barrier to nucleation. We have
\begin{eqnarray}
\Delta F^*
 &=&
\Delta F_c +4 \pi r_c^2 (m_c-m)^2\int_{r_c}^{\infty}
{\rm d}r\left[\frac{1}{2}\chi^{-1}
+\kappa\left(\frac{1}{r}+\frac{1}{\xi}\right)^2\right]
\exp\left[2(r_c-r)/\xi\right]\nonumber\\
&=&
\Delta F_c +4 \pi\kappa r_c^2 (m_c-m)^2\int_{r_c}^{\infty}
{\rm d}r\left[\frac{2}{\xi^2}+\frac{2}{\xi r}+\frac{1}{r^2}\right]
\exp\left[2(r_c-r)/\xi\right]
\nonumber\\
&=&
\Delta F_c +4\pi \kappa r_c(m_c-m)^2\left[1+r_c/\xi\right],
\label{fcrit}
\end{eqnarray}
where in obtaining the second line from the first
we substituted $\xi$ for $\chi$ using Eq. (\ref{xia}).
Finally, we can set $m=0$ to obtain the free energy barrier to nucleation
of the ordered phase of transition $\alpha$ near transition $\beta$,
\begin{equation}
\Delta F^*=
\Delta F_c +4\pi \kappa r_cm_c^2\left[1+r_c/\xi\right],
\label{fcrit0}
\end{equation}
As we approach transition $\beta$, $T\rightarrow T_{\beta}$,
$\Delta F^*$ approaches the finite limit
\begin{equation}
\Delta F^*(T=T_{\beta})= \Delta F_c +4\pi\kappa r_c m_c^2.
\label{fcp}
\end{equation}
The free energy $\Delta F^*$ can be written as
\begin{equation}
\Delta F^*=
\Delta F^*(T=T_{\beta})+A\left(\frac{r_c}{\xi}\right),
\label{fscale}
\end{equation}
where $A=4\pi\kappa r_cm_c^2$, a constant. The singular part of
$\Delta F^*$ has the form: the ratio $r_c/\xi$ raised to the power 1.

\subsection{Derivatives of $\Delta F^*$}

We can take the temperature derivative of $\Delta F^*$. As $m$ does
not vary with $T$ above $T_{\beta}$ we may use Eq. (\ref{fcrit0}),
and obtain
\begin{eqnarray}
\frac{\partial\Delta F^*}{\partial T} &=&
\frac{\partial\Delta F_c}{\partial T}
+4\pi\kappa r_c^2m_c^2\frac{\partial \xi^{-1}}{\partial T},
\label{dfdt}
\end{eqnarray}
which near $T_{\beta}$ becomes
\begin{eqnarray}
\frac{\partial}{\partial T}\left(\Delta F^*-\Delta F_c\right)
=
\left(2\kappa a\right)^{1/2}\pi r_c^2 m_c^2
\left(T-T_{\beta}\right)^{-1/2}
  ~~~~~~ T > T_{\beta},
\label{dtscale}
\end{eqnarray}
where we used Eq. (\ref{xi}) for $\xi$.
Just above the transition $\beta$ the derivative of the barrier diverges
to $+\infty$; the barrier drops very
rapidly with decreasing temperature just above $T_{\beta}$.

We can also take the derivative of $\Delta F^*$ with respect
to the field $h$ conjugate to $m$. 
Using Eq. (\ref{fcrit}), and taking note of the definition
of $\chi$, Eq. (\ref{chidef}),
\begin{equation}
\frac{\partial\Delta F^*}{\partial h} =
\frac{\partial\Delta F_c}{\partial h}
-8\pi\kappa r_cm_c\chi\left[1+r_c/\xi\right],
\label{dfdh}
\end{equation}
where after taking the derivative we set $m=0$. Note that
$\partial\xi/\partial h=0$.
As transition $\beta$ is approached the
rate of change of $\Delta F^*$ with respect to the field conjugate
to the order parameter diverges as the response function $\chi$.
Also, if we substitute
our solution for $m(r>r_c)$, Eq. (\ref{oz}), with $m=0$,
into Eq. (\ref{nstardef}), we obtain the size of the nucleus
\begin{equation}
\Delta m^*
= \Delta m^*_c
+8\pi\kappa r_cm_c\chi\left[1+r_c/\xi\right]
\label{ncrita}
\end{equation}
with
\begin{equation}
\Delta m_c=\int_{r\le r_c}
\Delta m({\bf r}) {\rm d}{\bf r},
\end{equation}
the contribution of the core, and we used Eq. (\ref{xia})
to substitute $\chi$ for $\xi^2$. Comparing
Eqs. (\ref{dfdh}) and (\ref{ncrita}), we see that
\begin{equation}
\frac{\partial}{\partial h}\left(\Delta F^*-\Delta F_c\right) =
-\left(\Delta m^*-\Delta m_c\right).
\label{nucth}
\end{equation}
For the fringe, the derivative of the free energy with respect
to $h$ is equal to minus the excess $m$.
This result is essentially what is called the nucleation theorem
\cite{kashchiev82,viisanen93,bowles00} in studies of nucleation
in fluids. It states that the larger the nucleus, the larger
$\Delta m^*$ is, the more rapidly the nucleation barrier varies
with $h$. In fluids $m$ is a number density difference and
$h$ is the chemical potential.

Returning to Eq. (\ref{ncrita}) for $\Delta m^*$ we see that although
the core can only contribute a finite amount
to $\Delta m^*$ as its volume is finite,
the contribution
of the fringe diverges
as transition $\beta$ is approached. The size of the nucleus
diverges as $\chi$ as the continuous transition is approached.
This result was first derived by the author in Ref. \cite{search}.
See also the earlier work of Talanquer and Oxtoby \cite{talanquer98}
who first suggested that the size of the nucleus diverges as a
critical point is approached. In Refs. \cite{search,talanquer98}
the critical point is that of a fluid-fluid or vapour-liquid--like
transition.

\section{Conclusion}

We have considered the effect of a continuous transition, transition
$\beta$, on the homogeneous nucleation of a new phase at a first-order
transition, transition $\alpha$. We found that the temperature derivative
of the free energy barrier to nucleation, $\Delta F^*$,
diverged as $(T-T_{\beta})^{-1/2}$ within our mean-field
theory, and that the size of the critical nucleus,
the nucleus at the top of the free energy barrier to nucleation,
diverged as the response function $\chi\sim(T-T_{\beta})^{-1}$.
The presence of a critical point makes the nucleus very
large, its diameter is the correlation length $\xi$,
and causes the free energy barrier to nucleation
to decrease rapidly with decreasing temperature.
It reduces the barrier and so facilitates nucleation.
This is just what was first demonstrated
by ten Wolde and Frenkel \cite{tenwolde97}
for nucleation of a crystalline phase near the critical point of
a fluid-fluid transition. It is a rather general phenomenon
and applies to any continuous transition with a scalar order
parameter, i.e., any Ising-like transition.
Whether or not the same effect appears near a continuous
transition in a system which is anisotropic or in which the order
parameter is not a simple scalar, is an interesting open question.

It is a pleasure to acknowledge discussions with R. Evans and A. Parry.
Work supported by EPSRC (GR/N36981).


\begin{thebibliography}{99}
\bibitem{note} Here we consider only homogeneous nucleation, nucleation
in the bulk, far from any interface. Heterogeneous nucleation, nucleation
at an interface, either with a wall or an impurity, is in fact more
common, and is not always an activated process.
\bibitem{critnote}
Note that conventionally `critical'
is used to denote both a continuous transition and the crystallite
at the top of the barrier although there is no connection between the
two uses of the word. Although this terminology is a little
unfortunate we will use it here and so we will be studying
a critical nucleus near a critical point.
\end{thebibliography}
\end{document}